# Post or Tweet: Lessons from a Study of Facebook and Twitter Usage

**Tasos Spiliotopoulos**

Madeira Interactive Technologies Institute

University of Madeira

Funchal, Portugal

tspiliot@gmail.com

**Ian Oakley**

School of Design and Human Engineering

UNIST

Ulsan, South Korea

ian.r.oakley@gmail.com

## Abstract

This workshop paper reports on an ongoing mixed-methods study on the two arguably most popular social network sites, Facebook and Twitter, for the same users. The overarching goal of the study is to shed light into the nuances of social media selection and cross-platform use by combining survey data about participants' motivations with usage data collected via API extraction. We describe the set-up of the study and focus our discussion on the challenges and insights relating to participant recruiting and data collection, handling and dimensionalizing usage data, and comparing usage data across sites.

## Author Keywords

Social Network Sites; Facebook; Twitter; Social media; media selection; API; behavioral data.

## ACM Classification Keywords

H5.3. Group and Organization Interfaces; Web-based interaction.

## Introduction

Social Network Sites (SNSs) are popular and diverse tools that present users with a wide range of features and options. The diversity of available services means that SNSs are often not used in isolation – many users adopt multiple services to fulfill their communication needs. Indeed, recent evidence indicates this behavior is becoming dominant – a 2013 survey indicates that 42% of online adults in the U.S. use multiple SNSs, more than the 36% who rely on just a single service [2]. In addition, there are substantial differences in the frequency and patterns of usage across services [2], suggesting that different SNSs are used to achieve different ends. There is even data indicating that at least one prominent user group (US teens) does not view Twitter (www.twitter.com) as a SNS and therefore may not associate it with more traditional services like Facebook (www.facebook.com) [9]. This reinforces the idea that different SNSs are used in very different ways and to satisfy different needs.

However, despite this diversity of offerings and use, most research on SNSs considers sites in isolation – for example there are many articles addressing Facebook or Twitter, but little literature that examines the same users' activity on the services together. This singular perspective on SNSs is highly problematic and may lead to confounded analyses. For example, consider a hypothetical group of Facebook users who post few textual status updates; half do not use Twitter while the other half tweet prolifically. We suggest that drawing conclusions about Facebook use from this group (for example, that they did not seek to share textual updates) would be highly misleading unless Twitter use were also captured.

This workshop paper reports on an ongoing mixed-methods study on the two arguably most popular SNSs, Facebook and Twitter, for the same users. In particular, the study examines user motivations for each service, and combines this information with usage data gathered computationally from the API for each participant. The overarching goal of the study is to shed light into the nuances of social media selection and cross-platform use. This paper describes the set-up of the study and focuses primarily on the challenges, findings and insights from the analysis of the computational usage data across platforms.

## Study Design

*Method*

Participants were recruited with a request to complete an online survey. Approximately 1/3 of participants were recruited through posts on social network sites, 1/3 through posts to online forums, mailing lists and online study repositories, and 1/3 through a Facebook ad campaign. The ad campaign consisted of two ads with similar wording targeted at self-reported English-speaking Facebook users from 12 countries. The participants were directed to a comprehensive study description page that clearly framed the experiment as an academic study and explained the data collection process. The participants then had to explicitly click a link to login with their Facebook credentials and access the survey, which is an equivalent action to installing a Facebook application. During this process the Facebook API ensures that the application displays all data-access permissions granted to it. Thus, we consider that the participants had a good understanding of the data captured by the study. Furthermore, participants had the option to opt out of the study at any time. After logging in, participants were directed to an online survey capturing demographics and their motivations for using Facebook (see [13] for discussion on the Uses and Gratifications theoretical framework behind eliciting these motivations for use). Then they were prompted to answer whether they were also Twitter users. If the answer was positive, they were presented with an additional set of questions capturing their motivations for using Twitter. In the background, the Facebook API collected a number of metrics about each participant's actual Facebook use. If they reported to be Twitter users and provided us with their (valid) username, we collected some public information about their Twitter account and usage. These data comprise an extended version of a previously described data set [13].

| Facebook usage metrics | Mean | SD |
|---|---|---|
| Posts made | 154 | 116 |
| Comments made | 96.7 | 145 |
| Likes (to posts, comments etc.) given | 338 | 543 |
| Photographs posted | 331 | 431 |
| Photograph albums created | 13.7 | 8.04 |
| Photographs tagged in | 84.7 | 252 |
| Activities mentioned | 13.3 | 34.5 |
| Likes (to pages) given | 181 | 320 |
| Check-ins made | 2.84 | 6.34 |
| Events attended | 1.37 | 2.25 |
| Facebook friends | 492 | 381 |
| Facebook groups joined | 22.7 | 31.6 |

Table 1: Facebook usage and network metrics collected

| Twitter account metrics | Mean | SD |
|---|---|---|
| Tweets | 1084 | 2456 |
| Followers | 169.1 | 362.9 |
| Friends | 238.4 | 416.3 |
| Followers' tweets | 2604 | 3136 |
| Followers' followers | 11241 | 36064 |
| Followers' friends | 4733 | 11223 |
| Friends' tweets | 5285 | 4147 |
| Friends' followers | 639592 | 846676 |
| Friends' friends | 6668 | 12497 |

Table 2: Twitter account metrics collected. The second-level metrics, i.e. those relating to a participant's friends and followers, are the average values for the friends and followers of each participant.

*Participants*
There were a total of 232 usable responses. The participants were 126 males (54.3%) and 106 females (45.7%), with a mean age of 24.9 years (SD = 8.68, median = 20, range = 14 – 62 years old). They came from 32 different countries, with 94 (40.5%) from the USA and 70 (30.2%) from India. The majority of the sample (75%, n = 174) were full-time students, 22% (n = 51) were employed, and 3% (n = 7) unemployed. On the days that they use Facebook, participants reported spending a mean of 78 minutes on the site. Out of the 232 participants in the study, 103 (44.4%) reported using Twitter. On the days that they use Twitter, participants reported spending a mean of 29.1 minutes on the service.

*Facebook and Twitter usage data*
The Facebook API was used to access a range of usage information for each participant in the form of 12 user activity variables. Out of the 103 participants that reported to be Twitter users, due to a collection miscalculation on our part we sought the data of only 82. Out of these 82 users, 13 opted not to give their Twitter username, and thus their Twitter data were not collected. Due to limitations of the Twitter API, the extended Twitter data (i.e., information on followers and friends) of further 11 users out of the remaining 69 were not collected. Tables 1 and 2 present descriptive statistics from the Facebook and Twitter usage data, respectively.

*Analysis Plan*
In order to explore the interplay between Twitter use and Facebook use we focus on whether and how Twitter usage and motivations affect Facebook usage and motivations. For this, we follow a four-step analysis. First, we conduct exploratory factor analysis on the Facebook usage data to identify dimensions of usage. Second, we conduct a binary logistic regression comparing Twitter non-users against all users in our sample, to identify high-level differences between these two groups (using the Facebook usage data as predictors). Next, we identify the motives for Facebook and Twitter use by conducting an additional exploratory factor analysis on our survey questions. Finally, for a more nuanced understanding of the motives mechanism, we hypothesize a model of relationships between the Facebook and Twitter motivations and conduct path analysis. The current workshop paper focuses on the first two steps of this analysis plan.

## Discussion

*Recruiting and data collection*
Usage of social network sites, and Facebook in particular, has most commonly been captured by self-report methods using surveys. Only recently researchers have identified the need to move away from these self-reported measures in favor of computationally collected usage data. Junco [3] found significant discrepancies between self-reported and actual Facebook use, while network scholars have struggled with issues such as recall bias [1], interviewer effects [10], and other sources of

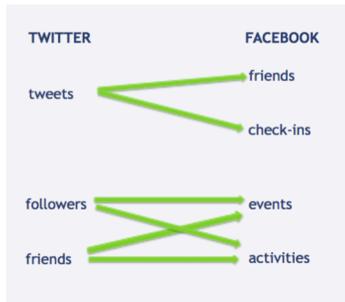

Figure 1: Cross-platform usage metrics correlations significant at the .001 level (out of a total of 108 tests). All six binary correlations are positive.

**Text communication**
- Comments made
- Likes (to posts, comments etc.) given
- Posts made

**Photographs**
- Photographs posted
- Photograph albums created

**Profile**
- Activities mentioned
- Likes (to pages) given

**Offline**
- Check-ins made
- Events attended

**Network**
- Facebook friends
- Facebook groups joined

Table 3: Summary of factors describing Facebook usage dimensions. One item, number of tagged photographs, loaded highly on two factors and was excluded.

measurement error that may accompany survey research (see [6] and [16] for discussion). Focusing specifically on information disclosure behavior, researchers have verified a discrepancy between stated privacy attitudes and actual behavior, termed "privacy paradox", and have suggested that this paradox can be overcome by studying people's behavior in realistic situations instead of lab experiments with self-reported behavioral data (e.g., [4]). More recent studies (e.g., [8,12,15]) have employed the Facebook API to gather broader and more granular data about users' online social activities.

Therefore, in this study we decided to complement users' self-reported activity on the sites with a range of data collected from the Facebook and Twitter APIs. Unpacking user activity into its constituents and taking advantage of the full wealth of data that can be collected programmatically was deemed more appropriate for a cross-platform study, because it would enable us to unearth specific nuances of use. This data collection process came with its own challenges, though. The online recruiting procedure led to a demographically skewed sample; a disproportionately large number of Indian participants were recruited due to the auction-like mechanics of Facebook ads. It is worth noting, however, that in the case of different, less exploratory studies such a recruiting approach could prove beneficial as it could facilitate the collection of a stratified sample. Further research should look more closely at identifying differences across such skews in the sample. In our case, for instance, users' nationality was found to have a significant effect on their motivations for using Facebook and had to be statistically controlled for [13]. Similarly, it is possible that the different recruiting methods, as well as features of each method such as the ad keywords used for targeting participants, can have a significant effect on sampling. Furthermore, even though we attempted to navigate around users' privacy concerns, it is apparent that our sample is subject to self-selection bias; not only participants self-selected to be included in the study, but they had to install a Facebook application and agree to offer some of their activity data through the API.

*Dimensionality of SNS usage*
The substantial breadth and scope of a site like Facebook often render overall descriptions of use too high level to be meaningful. However, blindly gathering and analyzing fine-grained usage data has its own conceptual and analytical pitfalls; treating each metric that can be collected from an API as an independent behavior may lead researchers to miss the larger picture of SNS usage and to misinterpret statistical findings. Following and extending previous work that demonstrates that information disclosure behaviors are multidimensional [4], we consider SNS *usage* as multidimensional and, thus, examine patterns of user behavior. Interpreting these dimensions of usage through their constituent items also allows a finer distinction across behaviors. We believe this approach will prove particularly useful in cross-site studies, which, for example, may strive to investigate differences among nominally similar behaviors (e.g., posting a photograph on Facebook and on Twitter).

Our data showcase these issues. For example, examining individual relationships among the 12 Facebook and 9 Twitter usage metrics would require conducting 12*9=108 correlation tests. Assuming an alpha threshold of .05, false positives are inevitable. In

| Measure | β |
|---|---|
| Age | -.034 |
| Gender (male) | .297 |
| Occupation (student) | -.505 |
| Nationality (USA) | -.138 |
| Text communication | .128 |
| Photographs | -.053 |
| Profile | -.199 |
| Offline | .567** |
| Network | .588** |
| Intercept | .861 |

*Nagelkerke $R^2$ = .161.*
*\* p < .05, \*\* p < .01, \*\*\* p < .001.*
*All beta coefficients are standardized.*

Table 4: Binary logistic regression predicting likelihood of a Facebook user being also a Twitter user.

our case, 22 tests were found to be significant at the .05 level and only six were significant at the .001 level (see Figure 1). It is worth noting that the numbers of a participant's Twitter friends and followers were very highly correlated (r=.946, p<.001), so some of the cross-platform correlations were due to this fact. Furthermore, it is interesting to note that none of the six "second-level" twitter metrics, that represent the activities of one's followers and friends, were found to be significantly associated with Facebook activity.

On the other hand, dimensionality reduction via factor analysis identifies five discrete dimensions of Facebook usage (Table 3) that can be clearly explained and interpreted, and can be used for further analysis (also see [14] for further discussion). This supports the argument that this analytical approach is both conceptually and statistically appropriate for this study.

*Comparing data across sites*
In addition to being statistically suspect, the correlational analysis shown in Figure 1 has the problem that it takes into account only a small subset of our sample - about one quarter. To address these problems, we focused initial analysis on whether or not a Facebook user is also a Twitter user and ignored the numerical data from Twitter. This allowed us to further reduce the number of variables and utilize our full sample to examine differences in Facebook usage between Twitter users and non-users. Of the five dimensions, only those that correspond to functionality not available in Twitter significantly (and positively) predicted ownership of an account, i.e. offline activity and Facebook network information (Table 4). This result indicates *complementary* use of the two SNSs based on feature differentiation [14]. This finding suggests an even more pronounced effect of feature differentiation, considering that previous research has found qualitative differences in the use of nominally similar features across platforms, e.g. linguistic differences between Facebook status updates and Tweets [7]. Further work in this area could focus on the temporal aspect of tandem usage, in order to identify specific user pathways.

We argue that this approach to analysis and presentation of the relationship between the Facebook and Twitter usage data provides more explanatory value and can help interpret the interplay between the usage of the two sites. Furthermore, while the current body of research on SNS non-use focuses on single sites to understand adoption and quitting (e.g., [5]), studying non-use in conjunction with usage of another site can significantly add to this body of work by providing much-needed context.

## Conclusion and ongoing work
In the context of an ongoing mixed-methods study of motivations and API usage data for Facebook and Twitter, this paper discussed challenges and insights relating to participant recruiting and data collection, handling and dimensionalizing usage data, and comparing usage data across sites. The overall goal of this research is to study cross-platform social media use through the lens of media selection; since motivations for use have been identified as main drivers of media selection [11], we expect that the combination of usage data with the motivations across SNSs can be useful for furthering the understanding of media selection processes in our ongoing work.